\documentclass[twocolumn,showpacs,aps,prb,superscriptaddress,floatfix]{revtex4}

\usepackage{graphicx}
\usepackage{amsmath}

\begin{document}

\title{Interaction Corrections to Two-Dimensional Hole Transport in Large
$r_{s}$ Limit}

\author{H. Noh}
\affiliation{Department of Electrical Engineering, Princeton University,
Princeton, New Jersey 08544}

\author{M. P. Lilly}
\affiliation{Sandia National Laboratories, Albuquerque, New Mexico 87185}

\author{D. C. Tsui}
\affiliation{Department of Electrical Engineering, Princeton University,
Princeton, New Jersey 08544}

\author{J. A. Simmons}
\affiliation{Sandia National Laboratories, Albuquerque, New Mexico 87185} 

\author{E. H. Hwang}
\author{S. Das Sarma}
\affiliation{Condensed Matter Theory Center, Department of Physics, 
University of Maryland, College Park, Maryland 20742}

\author{L. N. Pfeiffer}
\author{K. W. West}
\affiliation{Bell Labs, Lucent Technologies, Murray Hill, New Jersey 07974}

\date{\today}

\begin{abstract}
The metallic conductivity of dilute two-dimensional holes in a GaAs
HIGFET (Heterojunction Insulated-Gate Field-Effect Transistor)
with extremely high mobility and large $r_{s}$ is found to
have a linear dependence on temperature, consistent with the theory of
interaction corrections in the ballistic regime.
Phonon scattering contributions are negligible in the temperature
range of our interest, allowing comparison between our measured data
and theory without any phonon subtraction.
The magnitude of the Fermi liquid interaction parameter $F_{0}^{\sigma}$
determined from the experiment,
however, decreases with increasing $r_{s}$ for $r_{s}\agt22$, a behavior
unexpected from existing theoretical calculations valid for small $r_{s}$.
\end{abstract}
\maketitle

In two-dimensional (2D) charge carrier systems, it is well known that
any amount of disorder in the absence of interactions between the carriers
will localize the carriers, leading to an insulator with zero 
conductivity ($\sigma$) as the temperature ($T$) is decreased 
to zero\cite{loc}.
Recent experiments on high mobility dilute 2D systems, on the other
hand, have shown a ``metallic'' behavior at low $T$, 
characterized by an increasing
$\sigma$ with decreasing $T$, and an apparent metal-insulator 
transition (MIT) as the carrier density is lowered\cite{mit}.
There are three important energy scales in these systems. The first two
are the Fermi energy and the interaction energy. 
Their ratio, which is $r_{s}$, is around 10 or higher 
for the systems where the MIT is observed, implying that
interaction must be playing a role. 
The other energy scale is related to the disorder in the system
given by $\hbar/\tau$, where $\tau$ is the elastic scattering time.
It has been found from more recent experiments that disorder is also playing 
a significant role. In particular, the critical density ($n_{c}$ for 
electrons and $p_{c}$ for holes), above which a system shows the
metallic behavior, is found to decrease when disorder in the 2D
system is decreased\cite{yoon}.

An important question is whether this apparent metallic state
is truly a new ground state of the 2D charge carriers or 
simply a novel finite temperature behavior of the 2D gas,
since all experiments are done at finite $T$.
What is measured in such experiment is the temperature
coefficient, $d\sigma/dT$.
The metallic behavior evinced by the observation of negative $d\sigma/dT$
at finite $T$ does not necessarily mean, however, a true metal with nonzero 
conductivity at $T=0$.
Recently, Zala {\it et al.}\cite{zala} calculated the Fermi liquid interaction
corrections to the conductivity in 
the asymptotic low temperature regime ($T/T_F \ll 1$ where $T_F$ is
the Fermi temperature), and
pointed out that the metallic behavior seen
in the high mobility samples could be understood by taking into account
of interaction corrections in the ``high temperature'' ballistic regime
($k_{B}T\gg\hbar/\tau$).
They found that the conductivity of interacting 2D carriers changes linearly
with $T$ in the ballistic regime, $T_F \gg T \gg \hbar/k_B \tau$, 
with the sign as well as the
magnitude of $d\sigma/dT$ depending on the strength of the interaction,
while in the low temperature diffusive regime
($k_{B}T\ll\hbar/\tau$) the conventional logarithmically changing
conductivity\cite{int} is recovered.
A linear dependence of $\sigma$ on $T$ has also been predicted in earlier 
theories\cite{gold,das_sarma} based on temperature dependent
screening, and this screening contribution is included
in the theory by Zala {\it et al.}.

Experimentally, however, it is not straightforward to identify the interaction 
corrections unequivocally in the ballistic regime for two main reasons.
First, scattering by 
phonons can give significant contributions at high temperatures. 
In order to have the ballistic regime at sufficiently low $T$ to 
minimize the phonon contributions, the 2D charge carrier system must
have a very high mobility so that $\hbar/k_{B}\tau$ becomes very low.
Second, the temperature constraint, $T_F \gg T \gg \hbar/k_B
\tau$, satisfying the dual conditions of being in the ballistic
regime (i.e. $T\gg \hbar/k_B \tau$, which is a high-temperature
constraint) and of also being in the asymptotic low temperature
regime of $T \ll T_F$ (so that the thermal expansion in $T/T_F$,
essential for obtaining the linear-T term in the conductivity,
applies) is not easily satisfied experimentally, and indeed most
experimentally measured $\rho(T)$ data in 2D systems do not manifest
any clear cut linear-T behavior at low temperatures.
An additional issue we are addressing in this work is whether
the theory of interaction corrections to the conductivity can describe
2D transport in the large $r_{s}$ limit as well.
This is particularly germane in view of the fact that the
interaction theory is a systematic many-body diagrammatic expansion
in the interaction parameter $r_s$ (albeit an infinite order formal
expansion), and the question of the radius of convergence of
such an $r_s$-expansion becomes quite important for large $r_s$
values obtained in our samples.

In this paper, we report our experiments on the low temperature 
conductivity and the in-plane magnetoresistance (MR) of two-dimensional (2D) 
holes with extremely high mobility and very low density 
($r_{s}\approx$ 17 to 80) to study the interaction corrections.
From the temperature dependence of conductivity,
we clearly observed a temperature region where the the conductivity shows 
a linear dependence on $T$ even in this large $r_{s}$ limit
for a range of densities in the metallic side of the transition.
However, the Fermi liquid interaction parameter $F_{0}^{\sigma}$, determined
from a comparison of the data with the theory by Zala {\it et al.}\cite{zala}, 
shows a surprising non-monotonic dependence on the carrier density
with its value lying between $-0.5$ and $-0.7$.
$F_{0}^{\sigma}$ increases in magnitude with decreasing density for
$p\agt2\times10^{10}$ cm$^{-2}$ ($r_{s}\alt22$) and then
decreases with decreasing density for
$p\alt2\times10^{10}$ cm$^{-2}$ ($r_{s}\agt22$), a behavior
unexpected from a simple extrapolation of the predicted dependence 
of $F_{0}^{\sigma}$ on small $r_{s}$.
A separate measurement of effective g-factor ($g^{*}$)
from the in-plane MR provides a further confirmation of 
the unexpected behavior of $F_{0}^{\sigma}$.
$g^{*}$ decreases with decreasing density,
consistent with the behavior expected from $F_{0}^{\sigma}$ for
$p\alt2\times10^{10}$ cm$^{-2}$
through $g^{*}=g_{b}/(1+F_{0}^{\sigma})$.
 
The sample used in this study is 
a heterojunction insulated-gate field-effect transistor (HIGFET) 
made on a (100) surface of GaAs\cite{kane}.
A metallic gate, separated by an insulator (AlGaAs)
from the semiconducting GaAs, is used to induce the 2D holes at the
interface between the GaAs and AlGaAs. Ohmic contacts to the 2D holes are made
by using a self-aligned contact technique which allows the diffusion of the
contact material under the gate region. We would like to emphasize that
there is no intentional doping in the sample and the 2D holes are
induced by the applied gate voltage. 
This reduces the scattering by ionized
impurities so significantly that a very high mobility can be achieved.
The mobility ($\mu$) of the sample reaches $1.8\times10^{6}$ cm$^{2}$/Vs 
at a density ($p$) of $3.2\times10^{10}$ cm$^{-2}$, which is the highest 
achieved for 2D holes in this low density regime.
This high mobility makes $\hbar/k_{B}\tau$ for $p=3.2$ to $0.7\times10^{10}$ 
cm$^{-2}$ range from 16 mK to 80 mK, low enough that 
the temperature region where the metallic 
behavior is observed indeed corresponds to the ballistic regime while phonon 
contributions are negligible.
The extremely high mobility also allows us to measure the temperature
dependence of conductivity down to very low densities reaching
$p=1.5\times10^{9}$ cm$^{-2}$, with $r_{s}$ near 80 (assuming a hole
mass of $0.38m_e$).

\begin{figure}[t]
\begin{center}
\includegraphics[width=2.8in]{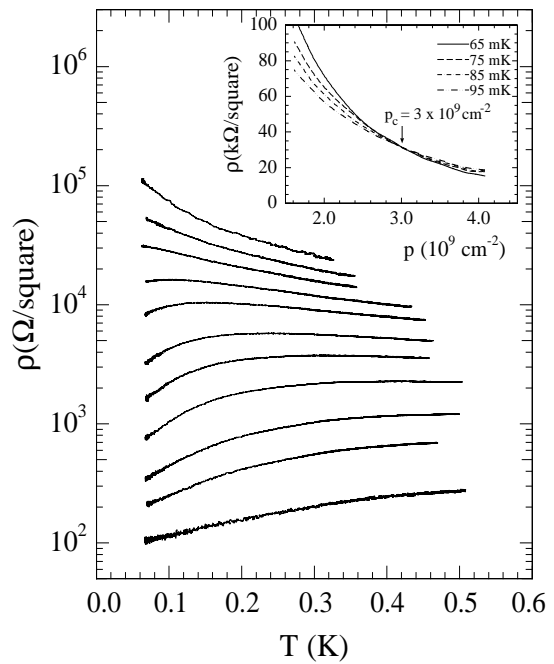}
\end{center}
\caption{\label{1}$\rho$ vs $T$ 
for $p=$ 3.2, 2.2, 1.7, 1.2, 0.9, 0.7, 0.5, 0.4,
0.3, 0.23, and 0.15$\times 10^{10}$ cm$^{-2}$ from the bottom.
The inset shows
$\rho$ measured as a function of $p$ at different $T$'s. 
The critical density $p_{c}$ is marked by an arrow corresponding
to a point where $\rho$ is temperature independent.}
\end{figure}

In Fig.~\ref{1}, we show the temperature dependence of the 
resistivity ($\rho$) at various densities. 
For $p\ge1.7\times10^{10}$ cm$^{-2}$, $\rho$ decreases 
monotonically with decreasing $T$, showing a metallic behavior.
For $p$ between $1.2$ and $0.4\times10^{10}$ cm$^{-2}$
$\rho$ shows a nonmonotonic dependence on $T$. It initially
increases with decreasing $T$ at high $T$, which was interpreted as
the classical to quantum crossover\cite{das_sarma}, and then decreases with
decreasing $T$ at low $T$ showing a metallic behavior
as the system goes into the degenerate regime. 
This crossover shifts to lower temperature with decreasing density
and the range where the metallic behavior is seen becomes very narrow,
especially for $p=0.5$ and $0.4\times10^{10}$ cm$^{-2}$.
For $p\le3\times10^{9}$ cm$^{-2}$, $\rho$ increases monotonically with
decreasing $T$, exhibiting an insulating behavior.
To identify the critical density, we measured $\rho$ as a function of $p$ 
at different temperatures and the data is shown in the inset.
The critical density determined from the crossing point, which shows a
temperature independent resistivity, is $p_{c}=3\times10^{9}$ cm$^{-2}$.
This low critical density, the lowest ever observed in 2D systems which 
exhibit the MIT, is
consistent with the previous
observation by Yoon {\it et al.}\cite{yoon} that the critical density becomes
lower with decreasing disorder in the system.
If we use the hole effective mass $m^{*}=0.38m_{e}$, this critical density
corresponds to $r_{s}=57$, which is much larger than the $r_{s}=37$
predicted for the Wigner crystallization in 2D\cite{wigner}.

\begin{figure}[t]
\begin{center}
\includegraphics[width=2.5in]{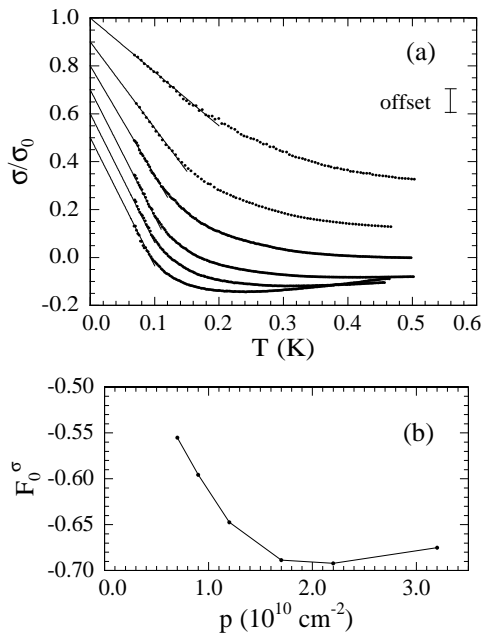}
\end{center}
\caption{\label{2}(a) $\sigma$ vs $T$ for $p=$ 3.2, 2.2, 1.7, 1.2,
0.9, and 0.7$\times10^{10}$ cm$^{-2}$ from the top.
The data are scaled by the value extrapolated to $T=0$,
and offset by 0.1 for clarity. The solid line for each curve is a linear fit
in the low temperature region of the data. 
(b) $F_{0}^{\sigma}$ vs $p$.}
\end{figure}

In Fig.~\ref{2}~(a), we replot the data for $p=3.2\times10^{10}$ 
cm$^{-2}$ to $0.7\times10^{10}$ cm$^{-2}$ as $\sigma$ vs $T$. 
The data are scaled by $\sigma_{0}$, the value of $\sigma$ extrapolated 
to $T=0$, and offset by 0.1 for clarity.
The metallic behavior is identified by increasing $\sigma$ with
decreasing $T$ at low $T$ for all these densities. 
Clearly, there is a region where $\sigma$ shows a linear dependence on 
$T$ as shown by the best fits in the figure
with solid lines. We note that $r_{s}$ for these densities ranges 
from 17 to 37 and $\sigma$ shows a linear dependence on $T$ for such
large $r_{s}$.
To compare our results with the theory by Zala {\it et al.}\cite{zala},
we need to consider several points.
First, as discussed below phonon contributions are negligible
in the temperature range where the linear dependence is observed,
which is below 200 mK for $p=3.2\times10^{10}$ cm$^{-2}$ and becomes lower 
for lower densities.
Second, this linear region is in the ballistic regime 
($k_{B}T>\hbar/\tau$) since
$\hbar/k_{B}\tau$ calculated from $\sigma_{0}$
ranges from 16 mK for $p=3.2\times10^{10}$ cm$^{-2}$
to 80 mK for $p=0.7\times10^{10}$ cm$^{-2}$.
Finally, this region is also much lower than the Fermi temperature ($T_{F}$) 
of the system, which is 500 mK for the lowest density $p=0.7\times10^{10}$ 
cm$^{-2}$. Zala {\it et al.} have pointed out that the regime where 
$F_{0}^{\sigma}$ can be treated as a momentum independent constant is 
$T\ll(1+F_{0}^{\sigma})^2T_{F}$.
A self-consistency check after we have determined $F_{0}^{\sigma}$ 
approximately (but somewhat weakly)
satisfies this condition.
All these allow a direct comparison of our data with their theory.
From their theory,
the slope of this linear dependence is directly related to the 
Fermi liquid interaction parameter $F_{0}^{\sigma}$ by the relation
\begin{equation}
Slope=\frac{m^{*}k_{B}}{\pi\hbar^{2}p}\left[1+\frac{3F_{0}^{\sigma}}
{(1+F_{0}^{\sigma})}\right].   
\end{equation}

Using $m^{*}=0.38m_{e}$ (which was obtained from cyclotron resonance of
high density ($p=5\times10^{11}$ cm$^{-2}$) 2D holes on (100) surface of 
GaAs\cite{stormer}), we have obtained from our data 
$F_{0}^{\sigma}$ as a function of $p$. The result is shown  
in Fig.~\ref{2}~(b). The value of $F_{0}^{\sigma}$ lies between 
$-0.5$ and $-0.7$ in the density range we measured from
$3.2\times10^{10}$ cm$^{-2}$ to $0.7\times10^{10}$ cm$^{-2}$.
For $p\agt2\times10^{10}$ cm$^{-2}$ ($r_{s}\alt22$), 
the magnitude increases with decreasing density.
Proskuryakov {\it et al.}\cite{pros} have also found that $F_{0}^{\sigma}$ 
for 2D holes in (311)A GaAs/AlGaAs heterostructure increases in magnitude 
with decreasing $p$ for $p=2$ to $8\times10^{10}$ cm$^{-2}$ with 
values between $-0.3$ and $-0.45$.
The change of $F_{0}^{\sigma}$ per density is similar in both 
experiments, while the magnitude of $F_{0}^{\sigma}$ in our measurement 
is much larger.
We note that $F_{0}^{\sigma}$ found from experiments on
2D electrons in Si-MOSFET\cite{silicon} also has much smaller 
magnitude, ranging from $-0.14$ to $-0.5$ for electron densities
1 to $40\times10^{11}$ cm$^{-2}$,
while the value found from p-SiGe by Coleridge {\it et al.}\cite{coleridge}
is somewhat comparable to ours, between $-0.55$ and $-0.65$.
What is surprising in our experiment, which explores the much lower density
(larger $r_{s}$) regime,
is that the magnitude of $F_{0}^{\sigma}$
does not increase monotonically with decreasing density.
When the density is decreased below $2\times10^{10}$ cm$^{-2}$ ($r_{s}\agt22$)
the magnitude of $F_{0}^{\sigma}$ decreases again. 
This is opposite to the predicted dependence of $F_{0}^{\sigma}$ on $r_{s}$
valid for small $r_{s}$. To our best knowledge, the dependence of 
$F_{0}^{\sigma}$ on $r_{s}$ when $r_{s}$ is large has not been
calculated theoretically, and it is not possible to compare our result with
any theoretical predictions at this time.

\begin{figure}[t]
\begin{center}
\includegraphics[width=2.5in]{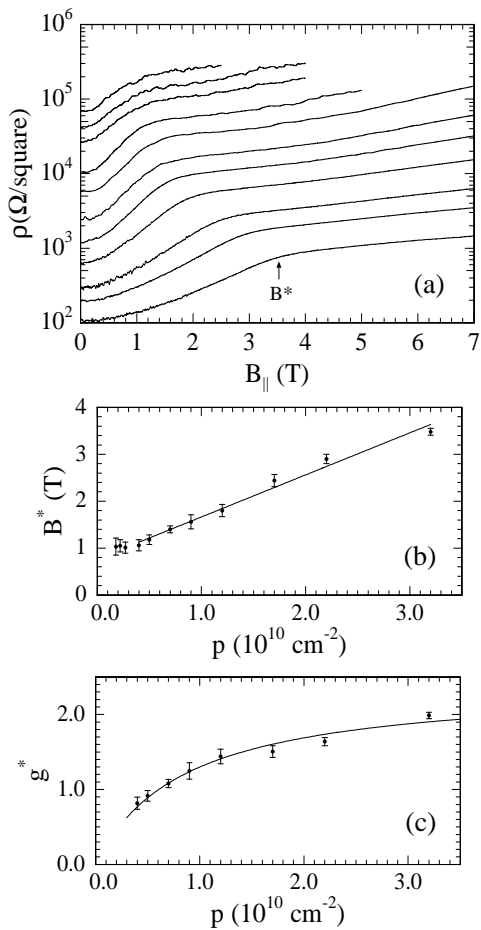}
\end{center}
\caption{\label{3}(a) $\rho$ vs $B_{\parallel}$ at $T=65$ mK and 
$p=$ 3.2, 2.2, 1.7, 1.2, 0.9, 0.7, 0.5, 0.4, 0.27, 0.22, and 
0.18$\times10^{10}$ cm$^{-2}$ from the bottom. 
$B^{*}$ is marked by an arrow for 
$p=3.2\times10^{10}$ cm$^{-2}$. (b) $B^{*}$ vs $p$. Solid line is a linear
fit to the data for metallic side, $p>p_{c}$. (c) $g^{*}$ 
determined from $B^{*}$. Solid line is the result obtained by the linear
fit in (b)}
\end{figure}

Experimentally, however, an additional test for this unexpected 
behavior of $F_{0}^{\sigma}$ can be made
from the MR measurements under an in-plane
magnetic field ($B_{\parallel}$). The in-plane MR provides a way to
measure the effective g-factor, which is directly related to 
$F_{0}^{\sigma}$ by the relation 
$g^{*}/g_{b}=1/(1+F_{0}^{\sigma})$, where $g_{b}$ is the bare g-factor. 
Figure~\ref{3}~(a) shows the in-plane MR measured in our sample 
for various densities at 65 mK. 
The MR increases as $exp(B_{\parallel}^{2})$ 
at low $B_{\parallel}$ and $exp(B_{\parallel})$ 
at high $B_{\parallel}$, consistent with an earlier observation for 
2D holes on (311)A GaAs\cite{yoon_b}.
Similarly strong MR has also been observed for 2D electrons in 
Si-MOSFET\cite{simonian} and the 2D electrons in GaAs\cite{tutuc}.
It has been established that this crossover from low field to 
high field dependences corresponds to full spin 
polarization of the carriers\cite{spin} and its position allows the
determination of $g^{*}$.
The crossover field $B^{*}$, determined from the position where the
second derivative of the $\rho$ vs $B_{\parallel}$ curve becomes 
maximum, is marked by an arrow for $p=3.2\times10^{10}$ cm$^{-2}$
in Fig.~\ref{3}~(a), and the dependence of $B^{*}$
on $p$ is shown in Fig.~\ref{3}~(b).
For $p>p_{c}$, $B^{*}$ decreases linearly with decreasing $p$ (best fit 
given by the solid line), and saturates in the insulating side of the MIT 
for $p<p_{c}$.
This behavior is also consistent with earlier observation by Yoon 
{\it et al.}\cite{yoon_b} for the 2D holes on (311)A.
A different way of determining $B^{*}$, using the inflection
point between high and low field dependences, yields a result within
the error bar of this plot, and produces an error of less than
15 \% in $g^{*}$.

For the metallic side, $g^{*}$ determined from the relation 
$2E_{F}=g^{*}\mu_{B}B^{*}$ 
(where $E_{F}$ is the Fermi energy and $\mu_{B}$ is the Bohr magneton)
is shown in Fig.~\ref{3}~(c) as a function of $p$.
$g^{*}$ decreases monotonically with decreasing $p$ for the density
range measured. 
The solid line in Fig~\ref{3}~(c) is the result when the best linear
fit in Fig~\ref{3}~(b) is used. 
Although a quantitative comparison between $g^{*}$ in Fig~\ref{3}~(c)
and the value of $g^{*}$ expected from $F_{0}^{\sigma}$ cannot be made since 
the bare g-factor $g_{b}$ is not well known for holes in GaAs,
we can qualitatively compare their density dependences.
From the density dependence of $F_{0}^{\sigma}$ shown 
in Fig.~\ref{2}~(b) and the relation
$g^{*}/g_{b}=1/(1+F_{0}^{\sigma})$, we expect that $g^{*}$
increases with decreasing $p$ for $p\agt2\times10^{10}$ cm$^{-2}$ and
then decreases with decreasing $p$ for $p<2\times10^{10}$ cm$^{-2}$.
The behavior of $g^{*}$ shown in Fig~\ref{3}~(c), therefore,
agrees well with that expected from $F_{0}^{\sigma}$ for
$p<2\times10^{10}$ cm$^{-2}$, where $F_{0}^{\sigma}$ decreases
in magnitude with decreasing $p$.
While this region of $p$ is where we have observed the unexpected density
dependence of $F_{0}^{\sigma}$, in this region there is good agreement 
in the behavior of $F_{0}^{\sigma}$ and $g^{*}$.
Thus, our in-plane MR measurements confirm the unexpected behavior
of $F_{0}^{\sigma}$ found from the $T$ dependence of $\sigma$.

The decrease in magnitude of $F_{0}^{\sigma}$ with decreasing $p$
(increasing $r_{s}$) is of great interest, and needs further 
examination. We, therefore, have also analyzed our data without
the assumption of a density independent mass $m^{*}=0.38m_{e}$. 
In this analysis, we used three independent relations 
(equation (1), $2E_{F}=g^{*}\mu_{B}B^{*}$, and
$g^{*}/g_{b}=1/(1+F_{0}^{\sigma})$)
between $F_{0}^{\sigma}$, $m^{*}$, and $g^{*}$ to calculate each quantity.
Since $g_{b}$ is not well known, we used it as a parameter, and found that
the density dependence of each quantity does not depend on a specific
value of $g_{b}$ while $g_{b}=0.5$ gives the best quantitative
agreement with $F_{0}^{\sigma}$ and $g^{*}$ determined earlier.
From this analysis,
$F_{0}^{\sigma}$ also exhibits a nonmonotonic dependence on $p$. 
The magnitude of $F_{0}^{\sigma}$ increases with decreasing $p$ 
for $p\agt1.5\times10^{10}$ cm$^{-2}$ and decreases with decreasing $p$
for $p\alt1.5\times10^{10}$ cm$^{-2}$. 
The decrease of $F_{0}^{\sigma}$ in magnitude with decreasing $p$
for large $r_{s}$ is still observed and confirmed 
once again.

Any explanation for this surprising result should take into account 
the large $r_{s}$ values in our system.
For $r_{s}\ge37$, the 2D system is expected to
be a pinned Wigner crystal. The critical density for MIT in our system 
corresponds to $r_{s}=57$, considerably larger than the critical 
$r_{s}$ predicted for this crystallization. 
The $r_{s}$ values for which we observed
the unexpected behavior of $F_{0}^{\sigma}$ range between 22 and 37,
close to that predicted for Wigner crystallization.
We note that there has been a Monte Carlo calculation\cite{kwon} of
Fermi-liquid parameters for $r_{s}$ up to 5, where $F_{0}^{\sigma}$
still increases monotonically in magnitude with increasing $r_{s}$.
To our knowledge, there is no theoretical calculation of the
dependence of $F_{0}^{\sigma}$ on $r_{s}$ when $r_{s}$ is large,
relevant to our experiment.
The question whether the crystallization is preceded by a 
ferromagnetic instability with $F_{0}^{\sigma}=-1$ has to be addressed 
as well.
From Monte Carlo calculations, Tanatar and Ceperley\cite{wigner}
have showed both possibilities of a diverging and a finite-valued spin 
susceptibility as $r_{s}$ increases toward the critical $r_{s}$ for
the Wigner crystallization.
Our result appears to imply that the ferromagnetic instability 
does not occur in the large $r_{s}$ regime of our 2D hole system. 

\begin{figure}[t]
\begin{center}
\includegraphics[width=2.8in]{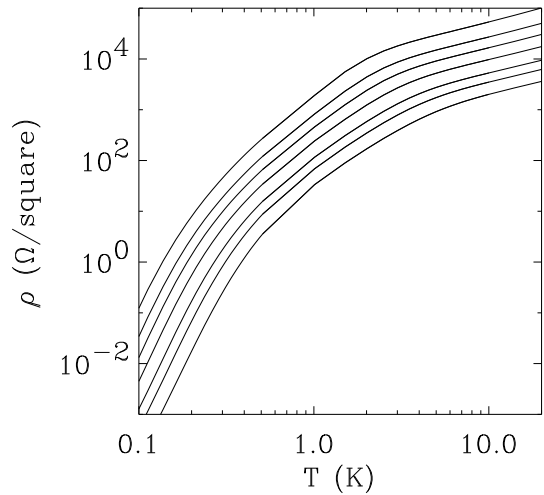}
\end{center}
\caption{\label{Fig_ph}Resistivity due to phonon scatterings (piezoelectric and
deformation potential coupling) as a function of temperature. 
Here the lines corresponds to the hole density $p=$ 0.5, 0.7, 0.9, 1.2,
1.7, 2.2, 3.2$\times10^{10}$ cm$^{-2}$ (from top to bottom).}
\end{figure}

We now address the important issue of phonon scattering
contribution to our measured hole resistivity, which we have ignored
in our analysis. The question of phonon contribution to the
resistivity is crucial since, if it is significant, it would then be
impossible to compare our measured resistivity to the interaction
theory. We have therefore theoretically directly calculated the
phonon scattering contribution by including both deformation
potential and piezoelectric coupling of the 2D holes to GaAs
acoustic phonons. Following ref. \onlinecite{kawamura} we have carried
out a detailed calculation of the phonon scattering contribution to
the hole resistivity in the parameter range of our experiment with
the results being shown in Fig. \ref{Fig_ph}. Our theoretical
phonon-only resistivity, as shown in Fig. \ref{Fig_ph}, demonstrates
that for $T\le 200 mK$, the temperature regime we concentrate on in
comparing our experimental resistivity with the interaction theory
(see Fig. \ref{2}), the phonon contribution to the resistivity is
miniscule (less than 1\% of the measured resistivity). We are
therefore justified in neglecting phonon scattering effects in the
discussion of our experimental results as long as we restrict
ourselves to $T < 200 mK$ as we have done in analyzing our data. As
is obvious from our theoretical results presented in
Fig. \ref{Fig_ph}, hole-phonon scattering contribution to hole
resistivity becomes non-negligible for $T>200mK$ and is, in fact,
significant for $T\ge 500 mK$ with its quantitative importance
increasing with decreasing carrier density.

We should point out in this context that we disagree with the
methodology employed recently by Proskuryakov {\it et al.} \cite{pros} in
subtracting out a calculated phonon scattering contribution to their
measured resistivity in analyzing their hole transport data in the
context of a quantitative comparison with the interaction theory of
Zala {\it et al.} \cite{zala}. First, it is well-known \cite{stern} that 
Matthiessen's rule does not apply to 2D systems at finite
temperatures, and therefore, subtraction of phonon contribution (even
if this contribution were accurately known, which is questionable) in
order to obtain the non-phonon part is unjustified and may be subject
to large errors. (This problem is worse in the presence of screening
of hole-phonon interaction, which must be included in the theory.)
Second, the calculation of phonon scattering contribution to hole
resistivity, following ref. \onlinecite{karpus}, carried out by
Proskuryakov {\it et al.} \cite{pros} is rather crude and approximate
(compared, for example, with our theoretical calculations shown in
Fig. \ref{Fig_ph} of this paper). We note that in
ref. \onlinecite{pros} the measured hole resistivity (before any
phonon subtraction) hardly manifests any clear cut linear temperature
regime, and the subtracted phonon contribution is a large fraction of
the measured resistivity, thereby casting substantial doubt on the
accuracy of the subtracted resistivity eventually compared with the
interaction theory. Our analysis in this work avoids these serious
pitfalls of ref. \onlinecite{pros} by directly considering the
measured resistivity in the context of interaction theory, which we
justify by explicitly calculating the phonon contribution to the hole
resistivity in the temperature range of our interest and showing it to
be negligible so that no arbitrary and unjustifiable phonon
subtraction is required (in contrast to ref. \onlinecite{pros}).

We calculate the temperature dependence of the hole resistivity by
considering screened acoustic-phonon scattering.
We include both deformation potential
and piezoelectric coupling of the 2D holes to 3D acoustic phonons of
GaAs. Details of the acoustic-phonon 
scattering theory are given in Ref. \onlinecite{kawamura}.
In this calculation we use the parameters corresponding to GaAs:
$c_l=5.14 \times 10^5$ 
cm/s, $c_t = 3.04 \times 10^5$ cm/s, $\rho=5.3$ g/cm$^3$, $eh_{14}=1.2
\times 10^7$ eV/cm, and $D=-8.0$ eV.
In the low temperature
range ($T<0.2K$) we find $\rho(T) \propto T^5-T^7$ because
the deformation potential scattering dominates over
piezoelectric coupling. In the intermediate temperature range ($0.2K < T
<T_{\rm BG}$) we have $\rho(T) \propto T^3$ mostly due to the
piezoelectric scattering \cite{karpus}. 
Above the Bloch-Gr\"uneisen temperature,
$T_{\rm BG}= 2 k_F c_l/k_B \approx 1-2K$, both scattering processes
give rise to linear temperature
dependence of the resistivity, $\rho(T) \propto T$.
The phonon contribution to the resistivity shows very weak
temperature dependence and is negligible when the
temperature is substantially below $T_{\rm BG}$. 
We emphasize that the phonon contribution to the resistivity cannot be
linear for $T<T_{\rm BG}$, and for $T \le 200 mK$ the phonon
contribution is negligible.

In summary, we have measured the temperature dependence of
the metallic conductivity of extremely high mobility
dilute 2D holes in GaAs in large $r_{s}$ limit. 
We find that the conductivity
shows a linear dependence on temperature in the ballistic regime.
The Fermi liquid interaction parameter $F_{0}^{\sigma}$ 
obtained from our data is found to 
exhibit a nonmonotonic dependence on density and decrease in magnitude
with increasing $r_{s}$ for $r_{s}\agt22$.

This work is supported by the NSF and the MRSEC at Princeton
University, and by US-ONR at Maryland.

\end{document}